\documentclass[oribibl]{llncs}
\usepackage[dvips]{graphicx}
\usepackage{makeidx}
\begin{document}

%%%%%%%%%%%%%%%%%%%%%%%%%%%%%%%%%%%%%%%%%%%%%%%%%%%%%%%%%
\mainmatter
\title{Exploiting Grids for applications in  Condensed Matter Physics} 
\author {Bhalchandra S. Pujari}
\institute {Department of Physics, 
         University of Pune, 
         Ganeshkhind, 
         Pune--411 007, 
         India.
\email{bspujari@physics.unipune.ernet.in} 
	 }
%$^2$Centre for Modelling and Simulations, University of Pune,  Ganeshkhind,
% Pune--411 007, India. 

%--------------------------------------------------

%%%%%%%%%%%%%%%%%%%%%%%%%%%%%%%%%%%%%%%%%%%%%%%%%%%%%%%%%
\maketitle

\begin{abstract}

  {\it Grids} - the collection of heterogeneous computers spread across the
  globe - present a new paradigm for the large scale problems in variety of
  fields. We discuss two representative cases in the area of condensed matter
  physics outlining the widespread applications of the Grids. Both the
  problems involve  calculations based on commonly used Density Functional
  Theory and hence can be considered to be of general interest.  We demonstrate
  the suitability of  Grids for the problems discussed and provide a general
  algorithm to implement and manage such large scale problems.

\end{abstract}

\section{Introduction}

The large scale computation has become an important tool in modern day sciences.
The applications of such calculations involve wide range of fields ranging from
atmospheric physics to quantum computing. {\sc Grids} offer a new dimension in existing
large scale computer infrastructure. Generally any large scale computation
involves  collection of machines which are aggregated in the form of clusters
and are located in vicinity of each other. On the other hand Grids are the
collection of several thousands of computers which are geographically separated by
large distances. Any given computing platforms may have heterogeneous architecture
and may be controlled locally by their own policies. Furthermore there may not exist
any dedicated networking backbone connecting each element of the Grid, thus
making the standard wired connection as the most widely used choice. Apart from its
heterogeneous components and wide spread locations Grid has a powerful application
porting system which takes care of each and every compute-job running on the Grid.
The so-called {\it middleware} accepts jobs from the user and assigns it to
different different computing nodes and at the end of the job the same middleware
returns the desired output back to the user. Such facility enables user to perform
the calculations without much of concern about explicit porting of any job.

As said earlier, the Grid consist a huge set of heterogeneous compute nodes, which
makes it an ideal tool for large number of jobs. Since the location of the nodes
are geographically far it is most suitable for non-parallel applications.
Thus, it is evident that such resource is most efficient if the given problem
can be split in several independent ones. With all this in mind we demonstrate here
how to handle the large scale problems  in condensed matter physics using the
power of Grids which are otherwise excessively expensive in terms of
time and CPU consumption.

Two problems discussed here involve the calculation of electronic structure which is used
frequently in condensed matter physics. The first problem involves the electronic
structure of quantum dots \cite{pujari,pujarici} while the second one deals with
evolution of atomic clusters \cite{kaware}. Both the problems are addressed using
commonly used Density Functional Theory (DFT). In the following section (Sec \ref{def})
  we discuss the general outline of the problems which also contains the computational
  details involved. In section \ref{suitable} we point out how the selection of the
  problems are  suitable for the Grids.  We present and discuss in brief the results
  obtained from our calculations in section \ref{results}.  It will be clear that the
  problems involve lot of compute jobs, the handling of which can become at time very
  painstaking.  We address this issue in section \ref{management} where we present the
  simple solutions for the management and implementation of such jobs. Finally
  conclude in Sec \ref{conc}.

\section{Definition of the problems\label{def}} 

In the following subsection we describe in details the nature of both
the problems and the computational procedure involved.  

\subsection{Quantum dots } 

Quantum dots \cite{reed,ashoori} are {\it zero dimensional} islands of electrons.
They are zero dimensional because the electrons inside the dots are under the
confinements from all three dimensions. In fact  quantum dots are the
manifestation of confinement of electrons by virtue of external potential. It is
quite similar to an atom, where the electrons are confined by Coulombic potential
($1/r$), except that, in the quantum dot the potential is tunable from outside.
Hence they are sometime called as {\it artificial atoms}.  

Applications of quantum dots range in a wide range of fields. From electronics to
biochemistry and from quantum computing to medical treatments. Apart from that,
being tunable in their properties, the quantum dots offer a playground for
physicists, both experimental as we theorists.  Experimentally the dots
manufactured in variety of ways like molecular beam epitaxy, electron beam
lithography, or self assembly via electrochemical means. No matter how they are
manufactured, the dots are always prone to some sort of impurities. To address
this issue,  we study a model impurity and its effects on the quantum dots.
Theoretically the quantum dots are investigated by various methods like density
functional theory (DFT),\cite{reimann} configuration interaction (CI),
\cite{pujarici} Quantum Monte Carlo (QMC),\cite{ghosal} Coupled Clusters method
(CC) \cite{ideh} and others. Out of which, DFT is easy to implement and proven to
be fairly accurate. In the present work we use spin density functional theory
(SDFT) which is later supported by CI method.

The confining external potential of the dot is modelled as 2D square well
potential and is given by \begin{equation} V_{ext} (x,y) \; =\; \left\{
  \begin{array}{cc} 0 & 0 \le x \le L; 0 \le y \le L \\ V_0 & {\textrm otherwise}
  \end{array} \right., \end{equation}.
  % and is shown in figure  \ref{square}.
Studied impurity is modeled using a gaussian potential given as :
\begin{equation}
  V_{imp}\; =\; A e^{-B(x^2+y^2)} \label{eq:imp}
\end{equation}

For any given number number of electrons the area of the dots is changed hence
changing the density parameter $r_s$ which is defined as: \[r_s \; =\; L
\sqrt{\frac{1}{\pi N}}\:, \] where $L$ is the length of the dot containing $N$
electrons. It is clear from this equation that for higher density, $r_s$ is
lower and vice versa.  In our calculations the barrier height $V_0$ is set to
1200 meV. The material of the dot is assumed to be GaAs. We also assume
effective mass approximation with an effective mass m$^*$=0.067 $m_e$, where
$m_e$ is the mass of an electron, and dielectric constant $\epsilon$ =12.9. The
units of length and energy are scaled to effective atomic units: effective Bohr
radius $a_B^*$ = 9.8 nm and effective hartree Ha$^*$=2Ry$^*$ =12 meV. In the
SDFT formalism, the Schr\"odinger equation in Kohn-Sham scheme reads as
\begin{equation} \left(- \frac{\hbar^2}{2m} \nabla^2 + V_{eff}^\sigma({\bf r})
\right)\psi_i^\sigma({\bf r})= \epsilon_i\psi_i^\sigma({\bf r}) \label{hamilt}
\end{equation}. The equation is solved iteratively where, in each iterations
the potential (or the density) is improved based on the feedback from earlier
iteration(s) till the input and output potentials (or densities) become
identical. The procedure is called as the self-consistency.   We use real-space
grid technique for the solution of Eq.  \ref{hamilt} For exchange-correlation
energy, we use the local density approximation.\cite{gori,tanater}

To summarise, the goal of this work is to understand the effects of impurity on
the quantum dots. To gain the better understanding,  up to  twenty-electrons
dots are considered with several sizes of the dot. According to DFT, there
exists a unique charge density for the given effective potential of the system
and vice versa, however {\it a priori} we do not know the effective potential
nor the density. Hence we have to guess for one of them.  Our technique
initiates the self consistency with one of the {\it several hundred} educated
guesses of charge density in search of energy minima, which assures the
detection of actual ground state of the system. As will be discussed in
subsequent sections, this problem involve running large number of {\it jobs} to
obtain the accurate results.

\subsection{Atomic Clusters}

The quest for equilibrium geometries of atomic clusters of Gallium - a work done
by Kaware {\it et al} \cite{kaware}  - is another example illustrating the
efficient use of Grid for condensed matter physics.\footnote{The work is carried
out in our lab and author is grateful to his colleagues for providing the data
prior to publication.} Similar work on larger scale for sodium clusters has been
carried out by Ghazi {\it et al} \cite{ghazi} who partially used Grids for their
work.

Atomic clusters are the aggregates of atoms. Understandably they are  the
building blocks for several nano materials. They are stable, bound and are
artificially created  (that is one of the reason,  they are different from a
molecule).  The main questions of interest are: If $N$ number of atoms come
together what kind of shapes they will form? How will that be different than
their {\it bulk} counterpart? What is the stability of such aggregate? Are they
reactive? What is the magnetic nature? And how is a nanostructure  built up
starting from single atom?  And so on.  Despite the large number of studies
\cite{baletto}, a clear evolutionary pattern  over a wide range of sizes has not
been developed. There is no clear answer to apparently simple question: how does
a cluster grow {\it atom-by-atom}?  To address these and many other questions
Kaware {\it et al} \cite{kaware} simulated a series of clusters containing 13 to
55 gallium (Ga) atoms.  They exhaustively study the growth of these clusters and
the study has revealed a peculiar {\it order-disorder-order} pattern.  Their extensive Density Functional calculations involve a
search of not only $\sim$ 40 ground state structures but also $\sim$ 5000
structures of isomers!  The shear extent of the problem demands a computationally
large scale infrastructure which is made available in the form of Grids.

\begin{figure}
  \begin{center}
    \includegraphics[width=8cm]{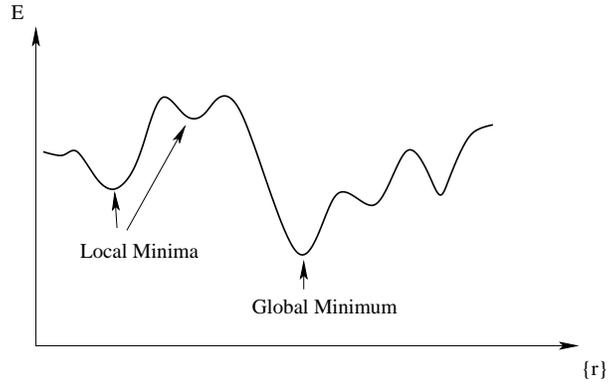}
  \end{center}
  \caption { A typical energy landscape depicting position of several
  local minima which indicate the isomers and a global minima which refers to
  ground state geometry. \label{landscape}}
\end{figure}

As stated earlier, the calculations are performed under Density Functional
framework within Generalized Gradient Approximation (GGA) \cite{gga}. The aim
of the simulation is to find out several equilibrium geometries (where the
forces on each atom are zero) and the lowest energy structure among those,
which is called as the ground state geometry.  Mathematically the energy $E$ of
a cluster is a function of potential $V(\vec r)$ which results due to
complicated interactions among the atoms.  
\[ 
E \; =\; \sum_{i < j} V_{ij}(\vec r_{ij}), 
\] 
where $i$ and $j$ are the indices associated with  atoms. This
gives rise to a typical energy landscape shown in figure \ref{landscape}. Each
minima on the landscape represents an isomer while the lowest of all minima -
called as global minimum - represents the ground state structure.

The procedure of finding the isomers is known as  {\it simulated annealing},
which involves non-linear optimization. Simulated annealing is the
theoretical  analogue of experimental technique, where the system
(cluster) is heated to a high temperature and then cooled down to obtain an
equilibrium geometry. If the system is slowly cooled then it is most likely to
reach its ground state geometry. On the other hand, if it is {\it quenched} it
reaches  one of its equilibrium geometries called {\it isomer}. 

Computationally, in simulated annealing, the cluster is heated (by providing
appropriate kinetic energy to the atoms) to a very  high temperature and then
quenched. This results in an equilibrium geometry. However to find out the
ground state, several hundreds of structures are required to be quenched.  In
other words, the problem is that of (non-linear) optimization {\it i.e.}, finding several
hundreds of equilibrium geometries, for $n$ interacting atoms. We need to do
this for $n$ ranging from, say, 10 to 50. Thus the total number of independent
executions, {\it i.e.}, total number of minimizations to be carried out, could
easily run into a few thousands, underlining the suitability of the Grid which
we shall see in the following section.

\section{Suitability of the grids \label{suitable}} 

In this section we illustrate the suitability the Grids for the given problems.  It is
clear from the discussion of earlier sections that both the problems involve large number
of jobs.  Here we quantitatively demonstrate that the number of jobs involve is too large
to run all the jobs on standard local compute machines.  Let us first consider the case of
quantum dots. As stated earlier, number of electrons in a given dot range from 2 to 20.
To bring out the effect of $r_s$ the width of the dot is also varied in five steps. As the
calculation is spin polarized, for any given number of total electrons  the number of
constituent up and down electrons are also varied.  More importantly, for each system ({\it
    i.e.} fixed number of up and down electrons, fixed width (and $r_s$) of the dot) it is
necessary to conduct several DFT runs (typically 100) with varying initial `guess' for
charge densities. Take an example of ten-electron quantum dot. There are five spin states
possible (from all ten electrons up to five up - five down). For five different widths of
the dot and 100 initial guesses there are about 2500 calculations to be performed! Further
similar set of calculations are to be done by adding the {\it impurity} potential
resulting in about $\sim$ 5000 jobs. Thus for all twenty electron quantum dot with
impurity problem involves tens of thousands of runs to be carried out in order to get the
results of desired accuracy. Although none of the jobs are CPU or memory intensive, it is
the shear {\it number} of jobs which make it difficult to perform the calculation on
simple compute system.

Similarly, enormous amount of calculations are involved in second problem. A typical
calculation involve the search for the ground state of a series of clusters involving at
least 10 clusters. Each cluster need several hundred initial geometries to be quenched.
The calculation also involve  repetition of  runs for  charged clusters (typically 2
    charged states). Thus, if we take 400 initial geometries then the total number of runs
of a series containing 40 clusters become : $40 \times 400 \times 2 = 32000$.

At this end we summarize the nature of the problems:
\begin{itemize}
\item Both the problems involve several runs
	\begin{itemize}
		\item Hundreds of initial guesses required for Quantum dots
		\item Hundreds and thousands of geometries to be quenched for clusters
	\end{itemize}
\item Each run is independent of the other.
\item None of the calculations require any specialized hardware 
\item and none require any specific need for parallelism
\end{itemize}
Thus, as can be understood the peculiarities associated with the problem make
them  extremely ideal to be implement on Grids.

\section{Results and discussion\label{results}}

In this section we briefly demonstrate the results obtained for both the
problems.  Detailed results are out of scope of the current paper and  we
strongly encourage our readers to refer to our work  published
elsewhere. \cite{pujari,pujarici,kaware, ghazi} Below we divide the results in
two subsections as per the problems discussed.

\subsection{Quantum dots} 

\begin{figure}
  \begin{center}
     \includegraphics[width=6cm]{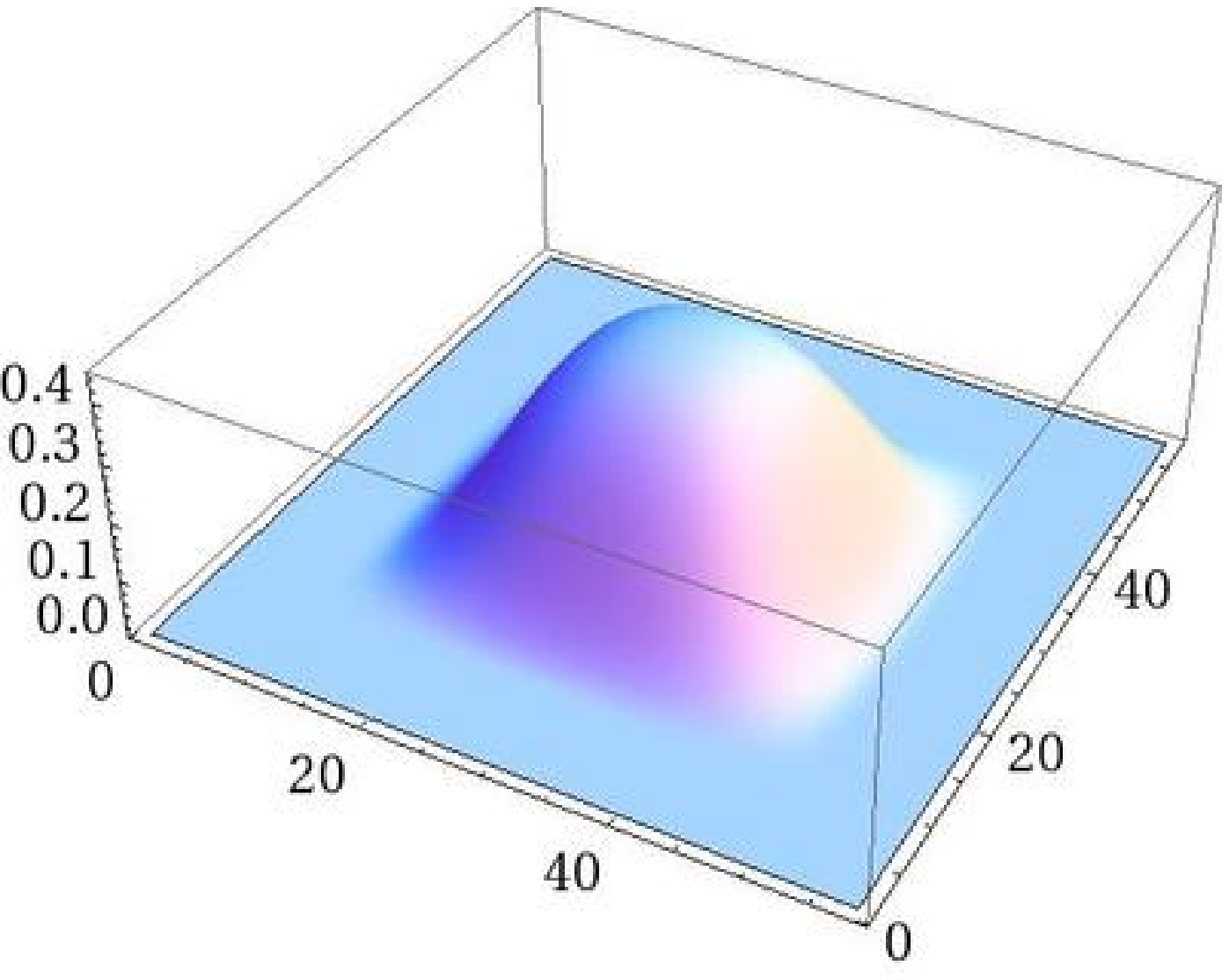}%\hskip 1cm 
     \includegraphics[width=6cm]{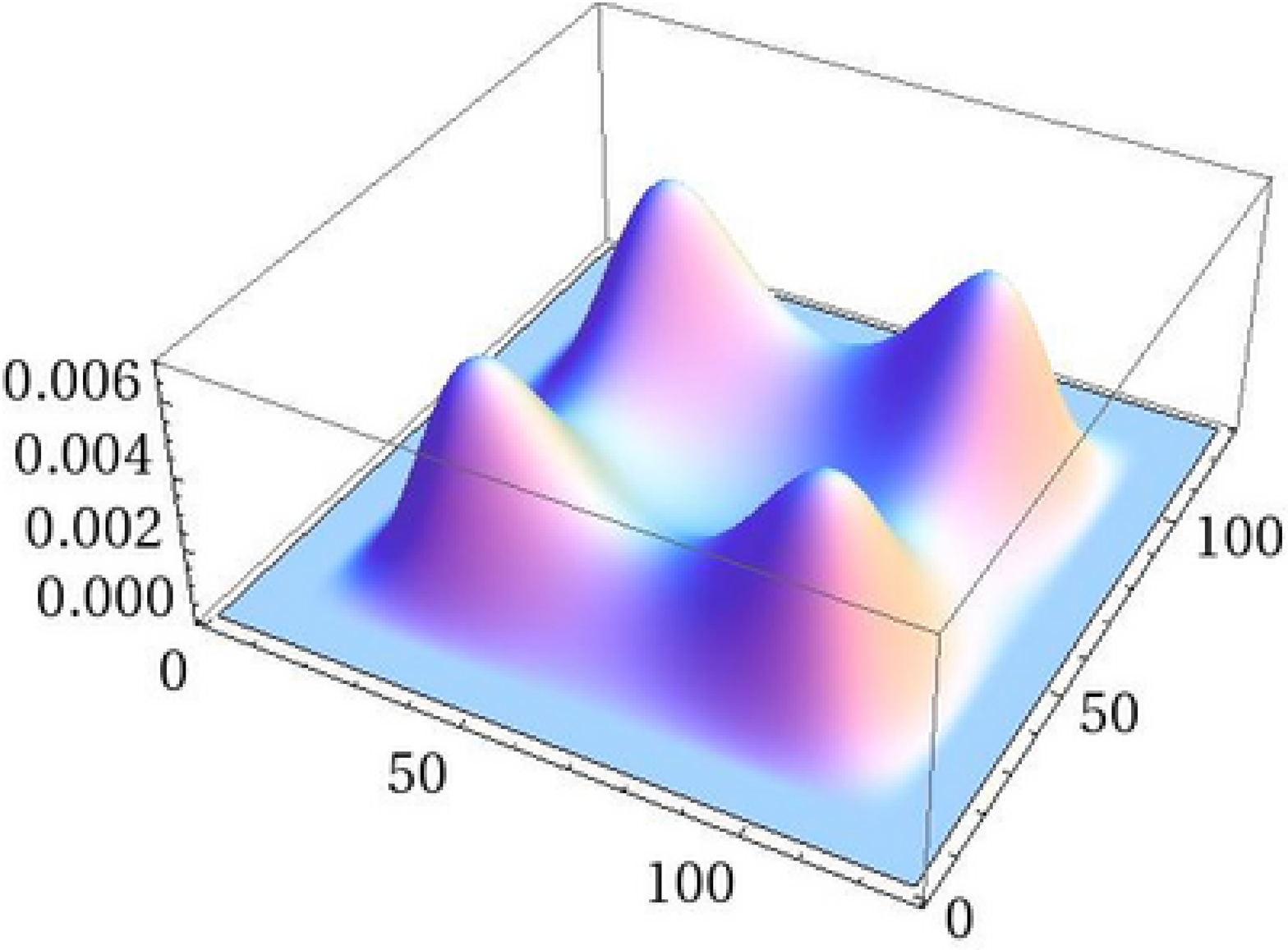}\\
     (a) \hskip 6 cm (b)
   \end{center}
  \caption{Typical electronic charge densities showing the feature of incipient Wigner
  localization for four electron quantum dot. (a) The charge density in low
  $r_s$ regime while (b) that in low density regime (high $r_s$). The imergence
  of four peaks in (b) is the indicative of incipient Wigner localization.\label{wigner}}
\end{figure}

We use density functional theory to investigate the quantum dots. One of
the major successes of DFT in quantum dots is to pick up a highly
correlated feature like {\it Wigner localization}. \cite{wigner} In such
confined electron systems, at low densities the confinement strength
weakens and the Coulomb interaction dominates over kinetic energy. As
the kinetic energy reduces the electron get localized to their
positions. Our calculations successfully pick up  a incipient Wigner
localization which is shown in the figure \ref{wigner}. Figure shows the
total charge densities of four-electrons quantum dots for two different
density regimes. The high density regime (small width of the dot) is
shown in figure \ref{wigner}(a) while low density regime is depicted in
(b). The emergence of four picks at the four corners is the typical
characteristics of the incipient Wigner localization. \cite{akbar}

\begin{figure}
  \begin{center}
     \includegraphics[width=6cm]{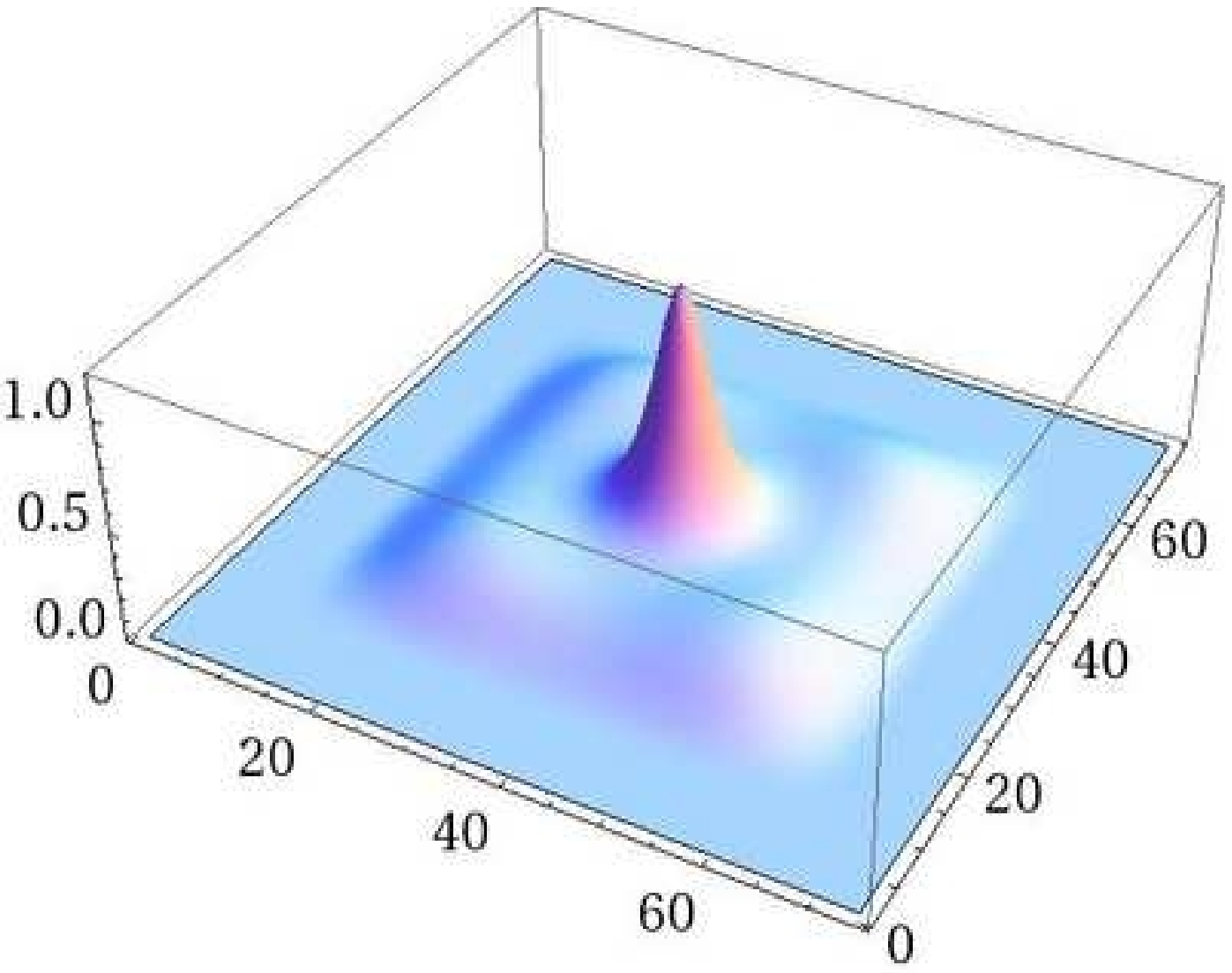} %\hskip 1cm 
     \includegraphics[width=6cm]{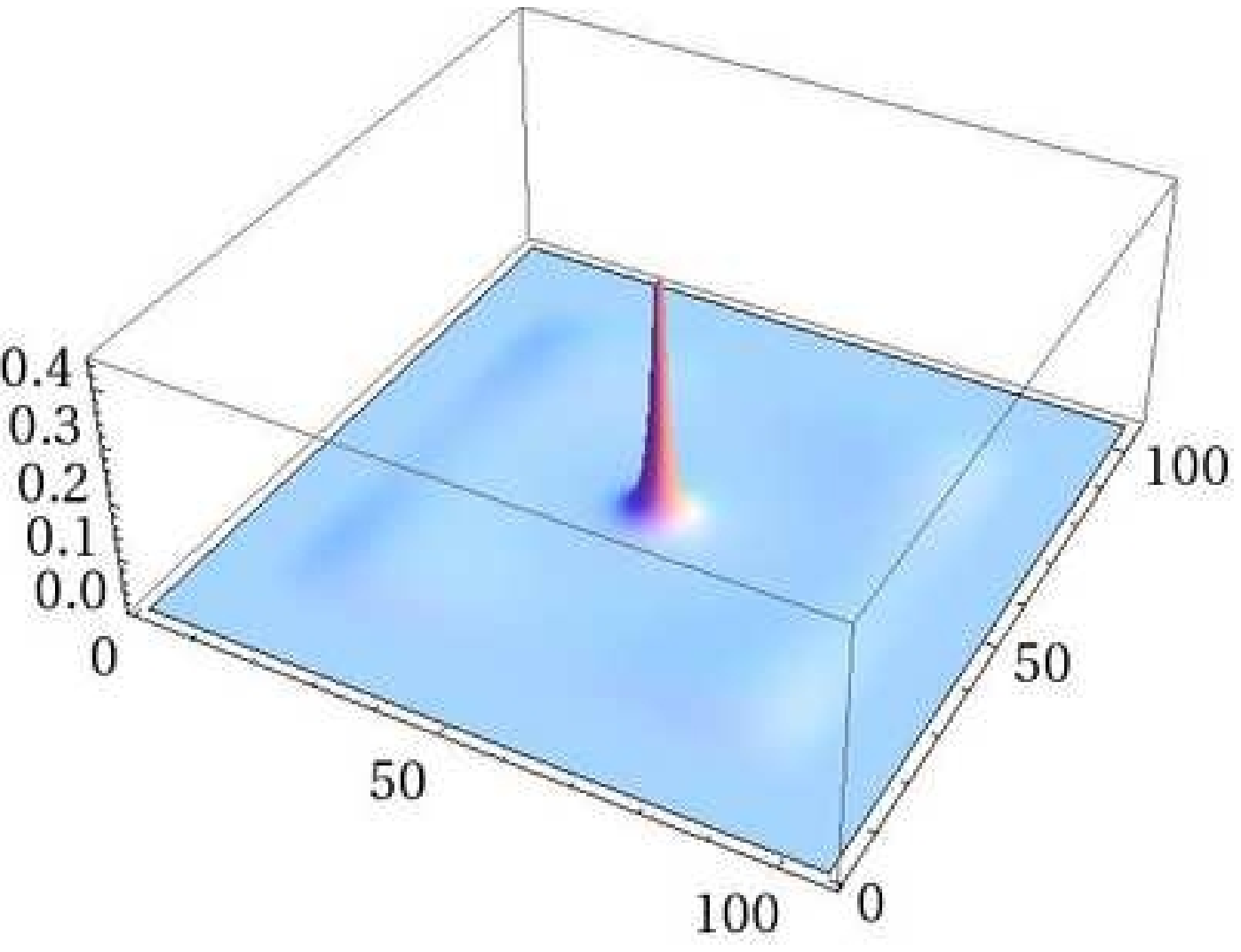}\\
     (a) \hskip 6 cm (b)
   \end{center}
  \caption{Evolution of the charge density of the dot as a function of
  dot size in presence of an attractive impurity. (a) Impurity being attractive
  in nature gives rise to peak seen at the center. (b) In low density
  regime the available area for electrons being sufficiently large the electron
  spread away and only electron trapped inside the impurity give relatively
  large peak. Four small peaks are also developed in the four corners.\label{impurity}}
\end{figure}

It is of equal interest to analyze the effect of impurity to on the
charge densities seen above. Figure \ref{impurity} shows the evolution
of charge density of same quantum dot in presence of the impurity.
Impurity being attractive in nature produces the peak in the charge
density. It should be pointed out that as the size of the dot is
increased the charge in the dot spreads over larger area while the
charge inside the impurity remain confined within the same region giving
rise to relatively large peak seen in figure \ref{impurity} (b).

The impurity is tuned in such a way that traps an electron inside it, thus
giving rise to localized magnetic moment. In many quantum dots this
localization is  associated with peculiar anti-ferromagnetic-like coupling with
firm unit magnetic moment at the center and four peaks at the corners for
opposite spins.  Our DFT analysis indicates that the presence of impurity may
change the ground state of quantum dot from magnetic to nonmagnetic and vice
versa. We also observe the oscillations in the charge density along the walls
of the dot as function of number of electrons.
%\section{Implementation and management}

\subsection{Atomic clusters}

\begin{figure}
  \begin{center}
  \includegraphics[width=4cm]{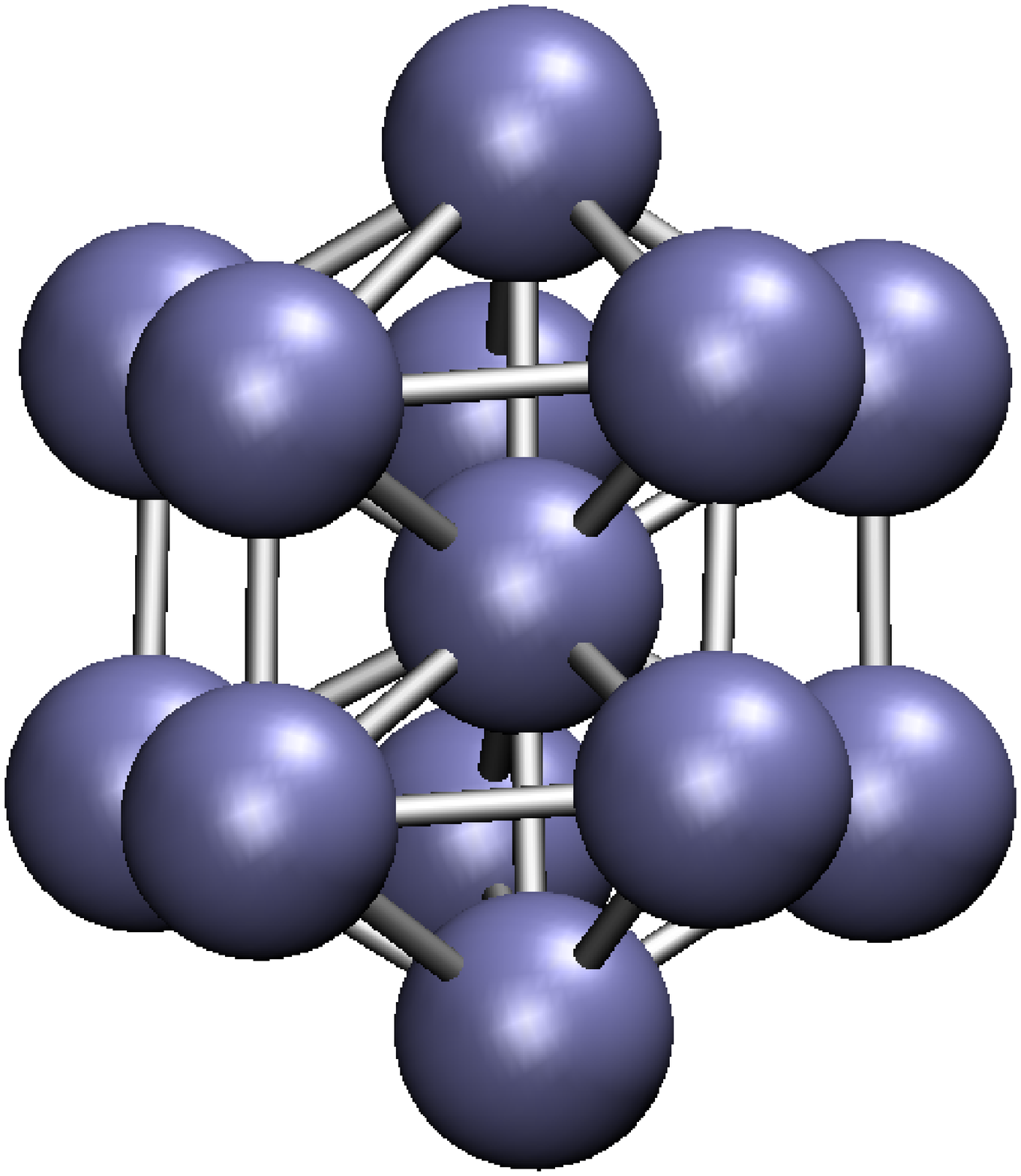} \hskip 1cm
  \includegraphics[width=5cm]{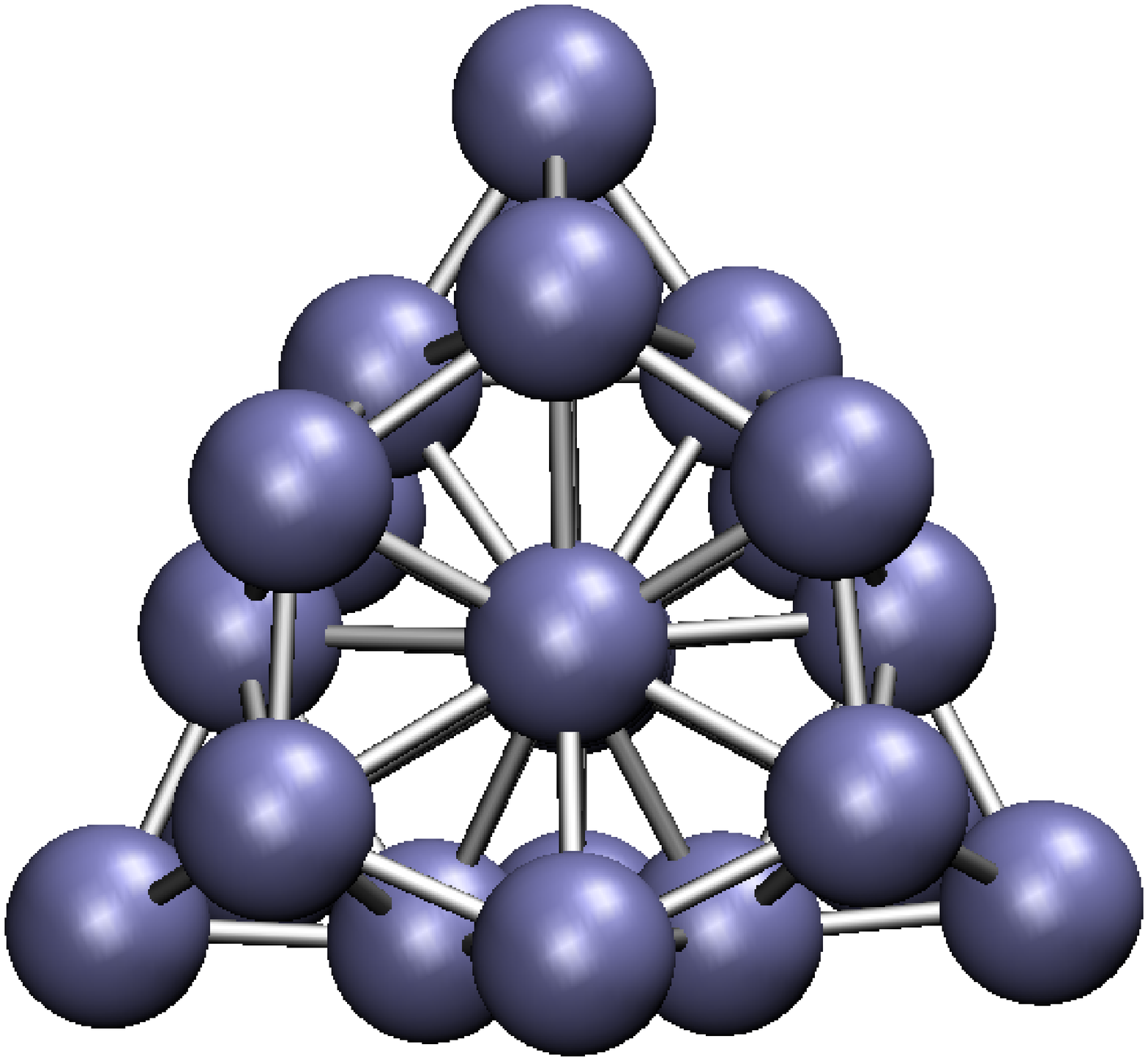}\\
  \centerline { (a) \hskip 5cm (b) }
  \includegraphics[width=5cm]{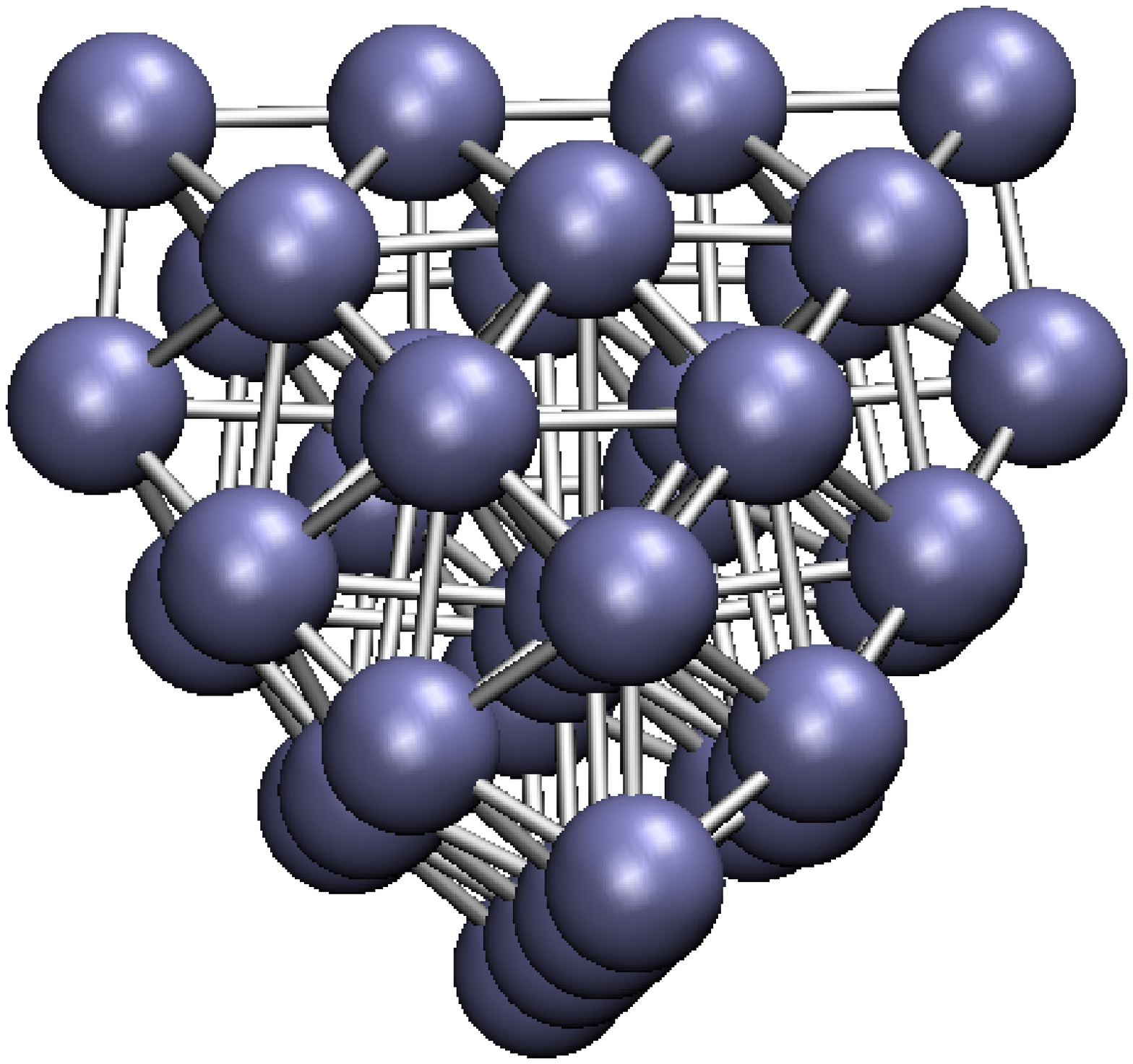} \hskip 1cm
  \includegraphics[width=5cm]{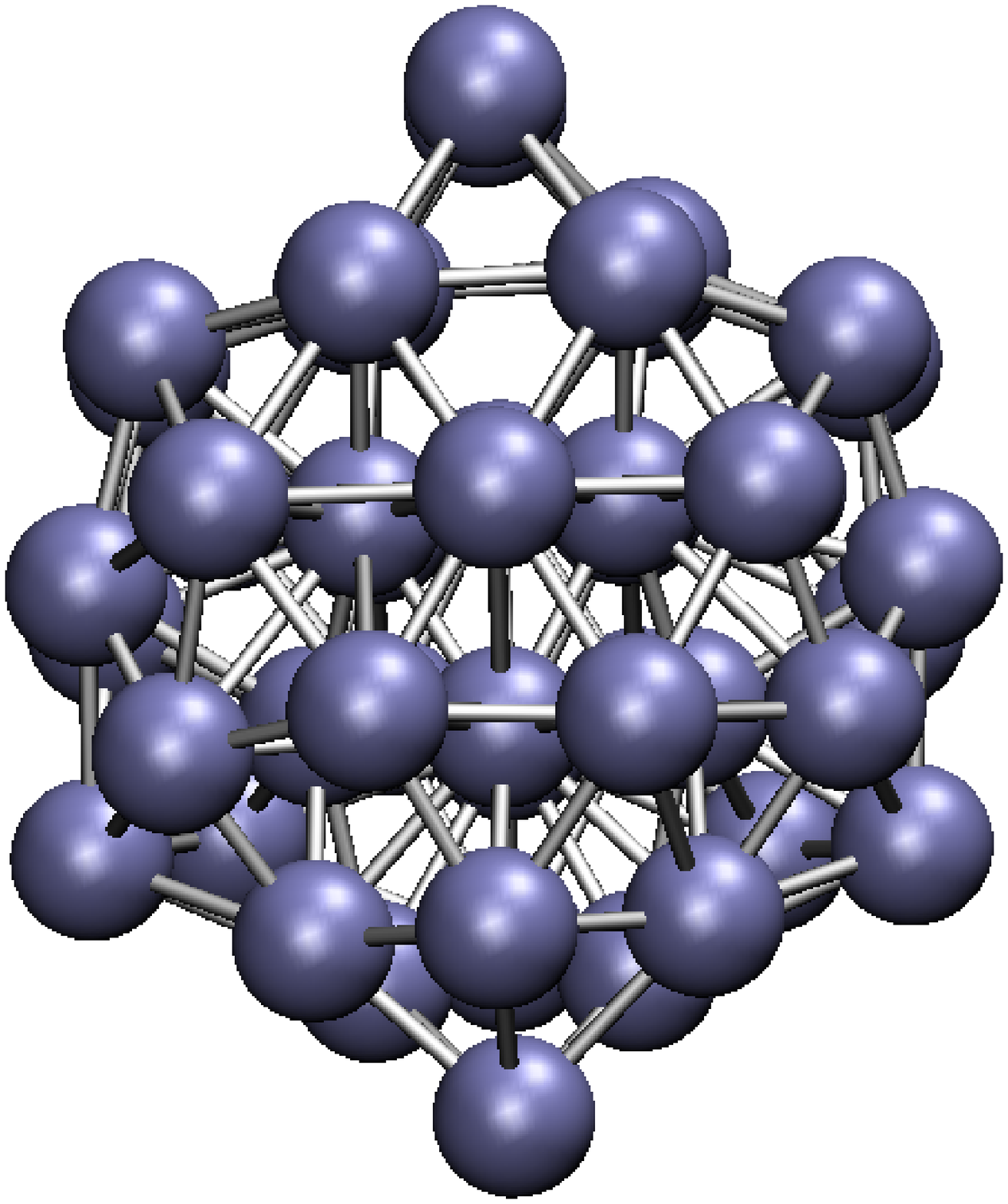}\\
  \centerline { (c) \hskip 5cm (d) }
   \end{center}
\caption{Evolution of the gallium cluster as we go on increasing the constituent
atoms. The geometries are for (a) Ga$_{13}$, (b) Ga$_{24}$,  (c) Ga$_{36}$ and (d)
Ga$_{47}$. \label{na:geom}}
\end{figure}

The main objective here is to obtains  several equilibrium geometries of gallium
clusters in the size range of  $n=$ 13-55 \cite{kaware}. 
Authors examined the evolutionary trends as
the clusters grow. Figure \ref{na:geom} shows few representative  equilibrium
geometries obtained, which highlight the evolution process of the shapes of
clusters with growth in their size. 

As can be seen from the figure, the geometries represent several ordered and
disordered structures. It was seen that addition of few atoms can drastically
change the order of the system. Similar observations  on larger scale were also
reported by Ghazi {\it et al} \cite{ghazi}. Gallium clusters show the tendency of
forming planer (or slab-like) structures. Further it was seen that most of the
bonds in the cluster are of sp$^2$ type, which is unlike aluminium clusters which
imply that the Gallium clusters do not fit into the simple {\it jellium}-like model.

\begin{figure}
  \begin{center}
    \includegraphics[width=8cm]{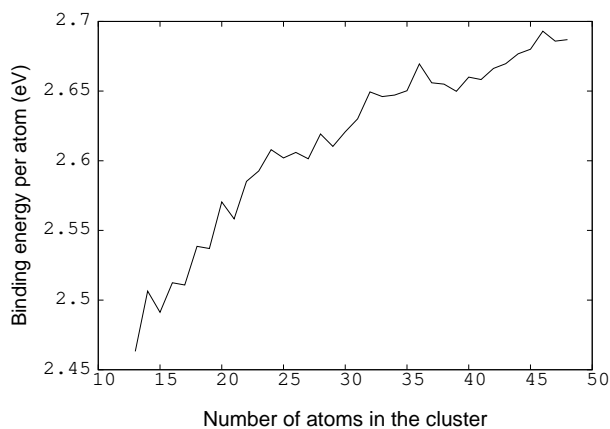}
  \end{center}
  \caption{Binding energy per atom for gallium clusters ranging from $n=13$ to
  48. Increasing binding energy per atom indicates that the larger clusters are more stable 
  than the  smaller ones.\label{be}}
\end{figure}

To examine the stability of the cluster it is instructive to analyze the binding
energy per atom of the cluster. Binding energy per atom is the amount of energy required to
remove an atom completely from the cluster. Thus, higher the binding energy
stronger the cluster. Figure \ref{be} shows the binding energy per atom for the
clusters ranging from 13 to 48. It is clear from the figure that the clusters with
increasing number of atoms are more stable. The binding energy per atom tend to saturate as
the number of atoms increases. 

Based on the conclusion of both the works \cite{kaware, ghazi}, it is clear that the growth shows an order-disorder-order pattern. In fact
we found that even in the disordered cluster there are hidden interlinked
ordered structures. Authors observed that between  two ordered structures the
growth proceeds via disordered clusters having multicentered icosahedral local
order. The transition from disordered to ordered structure is rather sharp and
occurs merely on changing the number of atoms by two or three. It was also found that the geometries strongly influence the melting temperature of the given cluster. 

\subsection{Management of the jobs\label{management}}

A typical problem faced when we handle such  large scale problems is the
implementation and the management of the jobs involved. All together we have
several thousands of jobs in both the problems and it is extremely desirable
to have a tool which can assist in handling such enormous number of jobs. Understandably
submitting, monitoring and retrieving each job manually is a tedious and
time consuming procedure and any web-based application may turn out to be
inefficient. We seek the simple solution in the form of {\it shell scripts}. It
turned out that the scripts are easy to use, highly customizable and equally
efficient tool for implementing and managing the jobs.

\section{Summery \label{conc}}

Thus to summarize, we have successfully implemented the Grids for the large
scale problem in the condensed matter physics. We have demonstrated that the
commonly used Density Functional Theory based calculations can be performed on
Grids. The nature of the problems involve large number of independent jobs to
be carried out where the Grid turned out to be most useful. For the management
of the jobs we mainly relied on standard Shell Script instead of any web-based
porting tool. 

\section*{Acknowledgments} 

Author like to thank D. G. Kanhere for valuable discussion,  Vaibhav Kaware,
Seyed Mohammad Ghazi, Kavita Joshi, Manisha Manerikar and Shahab Zorriasatein for
their contributions and Dr. Stefano Cozzini and Neeta Kshemkalyani for technical
assistance. It is a pleasure to acknowledge EU-India Grid project (Grant No.
RI-031834) for partial financial support as well as Garuda India Grid and C-DAC
for computing resources.

%\section*{References}

 \bibliographystyle{unsrt} % Bibliography style file, unsrt.bst
 \bibliography{biblio}

\end{document}